# The Grid: Prospects for Application in Metrology

P.I.Neyezhmakov – PhD, 1st Deputy General Director on scientific work of National Scientific Center "Institute of Metrology"( Kharkiv, Ukraine), Head of COOMET Secretariat (Euro-Asian cooperation of national metrological institutions)
S.I.Zub – senior scientist of NSC "Institute of Metrology", Kharkiv, Ukraine
S.S.Zub – PhD, doctoral candidate of Taras Shevchenko National University of Kyiv, Ukraine

## Abstract

Global system of distributing computing – Grid – created as reply for challenges, connected with the qualitative progress of complexity of experimental physical assemblies and information systems, is presented as optimal IT platform for assurance of measurement traceability in geographically remote regions and measurement data protection in global networks. The new component grid - Instrument Element (IE) – is intended for secure, remote, joint team work on monitoring and managing instruments generated and stored on distributed scientific equipment using conventional grid resources. The article describes the variety of all possible IE applications within grid technology for the tasks of metrology demanding IT support. Expanded by the new component IE grid becomes an optimal environment for effective monitoring, management and servicing of measuring resources which has the highest level of measurement data transfer, storage and processing safety and reveals new opportunities to track measurement procedures and assure a high level of confidence to these measurements.

## Introduction

International organisations such as OIML play an important role in the coordination of metrological activity in the globalised world [1].This activity is primarily intended to assure measurement traceability in geographically remote regions as well as to achieve a high level of confidence in the measurements.

The coordination and management of measurement processes often separated by long distances should be supported by state-of-the-art information technology. It is believed that this support could be based on the Internet. However, this system provides no possibility to control the measurement and computation processes, and the data are under constant threat of being falsified or even irretrievably lost.

The problems of interaction between remote devices as well as of measurement data protection in the global networks are being overcome this way or another [2,3,4], but the solutions are always partial, problem-dependent and take a lot of special effort.

Is there any way out? The answer is yes. Based on the same hardware (i.e. the same communication elements as used by the Internet) a *brand-new* system of globally distributed computations can exist assuring the highest possible level of safety for today which excludes such painful phenomena as viruses and hacker attacks.

This system appeared to meet the challenges of qualitative complexity of experimental physical assemblies and information systems which require new monitoring, management and operating approaches. The new paradigm of distributed computations is named Grid.

There are a number of implementations of this concept in the USA and Europe. However, the most wide-spread implementation of the Grid concept in recent years is the gLite middleware. It is the gLite that is used to support the most difficult physical installation created by mankind, the Large Hadron Collider (LHC). The development of this middleware is closely connected with the experiments conducted in the European Council for Nuclear Research (CERN) [12-15]. It is there where the most sophisticated measuring systems are now accumulated and it was there where the problem with the lack of computing resources to process the great amount of information obtained during LHC experiments appeared for the first time. However, despite the historical roots of the European DataGrid (EDG) project, and then the LHC Computing Grid

(LCG), the gLite middleware was developed as a *general platform* for constructing a distributed computing network [9,10,15]. With the financial support from the European Union the European scientific community launched the Enabling Grids for E-sciencE (EGEE) project within which the gLite middleware was created.

Remote control and data acquisition are part of the existing Grid concept. However, grid developments are more often focused on the sharing of computational and storage resources. In fact, these resources are much needed for most applications and users.

**Classical Grid Structure**

A classical grid is composed of the following basic structural elements (Figure 1): a set of computational nodes (PC) with installed Worker Nodes (WNs); a set of Resource Centres (RC) comprising the Computing Element (CE) and the Storage Element (SE); a set of basic Grid Services (the gLite middleware) with User Interfaces (UI) for access to Grid.

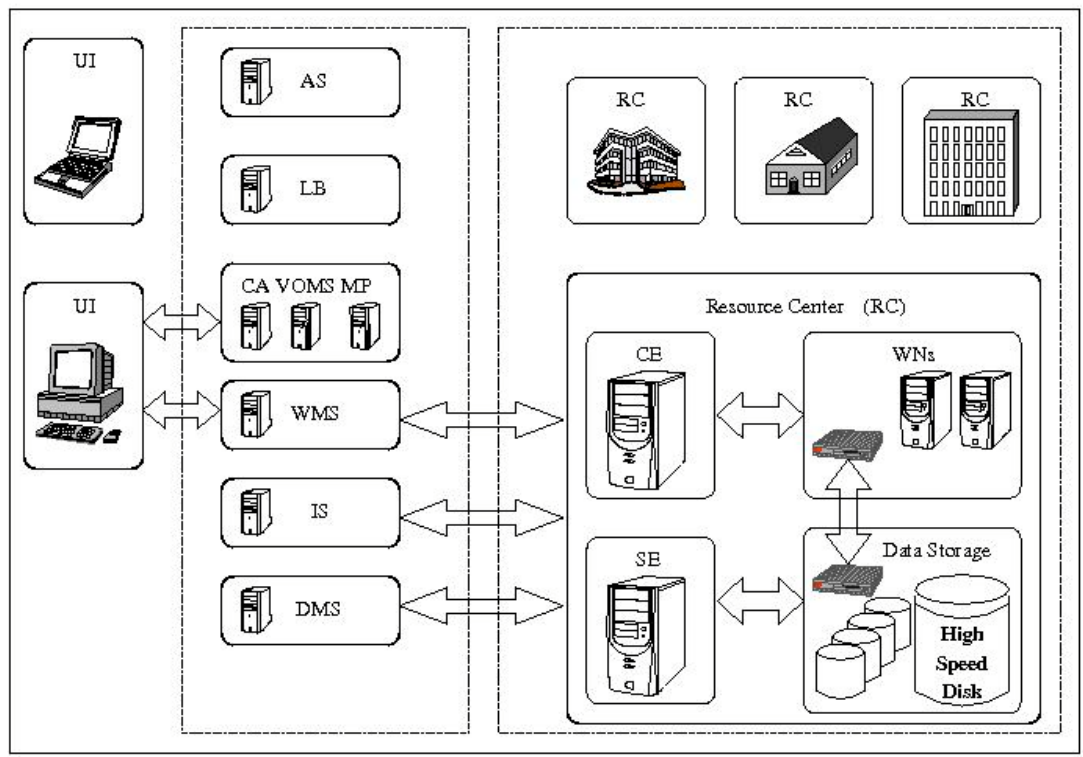

Figure 1. Grid system flow-chart.

The gLite package is a complete solution for the grid including both the basic low-level programs and a number of high-level services. It accumulates the components of the best projects existing today such as the Condor and Globus, as well as the components developed within the LCG project [15]. The gLite is compatible with such task planners as PBS [9], Condor and LSF, and contains the basic services facilitating the creation of grid applications for any applied area.

The Globus Toolkit developed by American scientists has actually become a world standard. It particularly includes a special HTTP-protocol for the use of the Grid Resource Allocation Management (GRAM); an extended version of the Grid File Transfer Protocol (GridFTP); the Grid Security Infrastructure (GSI); distributed data access based on the Lightweight Directory Access Protocol (LDAP); remote data access via GASS (Globus Access to Secondary Storage) interface.

A resource centre is composed of two types of resources [9,10]: computing resources controlled via the Computing Element (CE), and data storage resources accessible via the Storage Element (SE) which makes it possible to store and transport data between similar resources or between a resource and a grid user.

Basic Grid Services assure the operation of the entire grid system and are divided into the following parts: Workload Management System (WMS); Data Management System (DMS) consisting of a file directory service and a metadata directory service; information and monitoring system (IS) solving the problem of collecting and managing data on the status of the grid infrastructure; Grid Security Infrastructure (GSI) consisting of Certificate Authority (CA) issuing and supporting certificates, Virtual Organisation Membership Service (VOMS), MyProxy Service (MP) (the task of this part is to provide the authentication and authorisation between various grid system components); Logging and Bookkeeping (LB) System tracking task realisation processes; Accounting Subsystem (AS) designed to register the use of computing resources.

Thanks to the gLite the geographically distributed set of resources is represented as a single resource for users. *The gLite middleware plays the role of a personal computer operating system in the grid.*

## Security in the Grid

Security is at the heart of the Grid. Its components are both special secure file transfer protocols such as GSIFTP, Secure RFIO, Gsidcap, and an authorisation system which, in turn, is layered and deeply separates the grid resource access rights.

GSIFTP provides the functionality of the FTP protocol, however with support of the Grid Security Infrastructure (GSI). GSI assures the security in insecure public data networks giving such services as authentication, file transfer confidentiality and grid unified logon. GSI uses X.509 digital certificates as user and resource identifiers. This protocol is responsible for fast, safe and efficient file transfer. It makes it possible to control file transfer between two storage elements separated from the user (third-party transfers), as well as to transfer data in a few parallel flows.

What is new is the use of the Virtual Organisation Membership Service (VOMS) which stores information on the affiliation of users to certain groups and virtual organisations and their role in them. User rights can be set depending on user affiliation to a particular virtual organisation, group or role (with VOMS).

Thus, the grid security system has all known cryptography tools and a harmonious authorisation and right separation system based on electronic certificates.

## Instrument Element

However, there are more and more cases when besides resource distribution *close interaction between the instruments and grid users* should be assured. They often need to have a remote access to them, in principle, from any grid site.

The Grid Enabled Remote Instrumentation with Distributed Control and Computation (GridCC) project [17] was launched to meet these challenges. The project aims to use grid capabilities for secure, remote, joint team work on monitoring and managing instruments and data generated and stored on distributed scientific equipment using conventional grid resources.

The Instrument Element (IE) has been developed within this project which has been successfully used in various scientific collaborations for remote interaction with devices in the grid environment.

The IE consists of a linked service collection assuring the functionality for configuring, managing, monitoring of measuring instruments outside the IE interface which allows their interaction with the rest of the grid. Figure 2 illustrates the interaction between the IE and its users and other grid components [16].

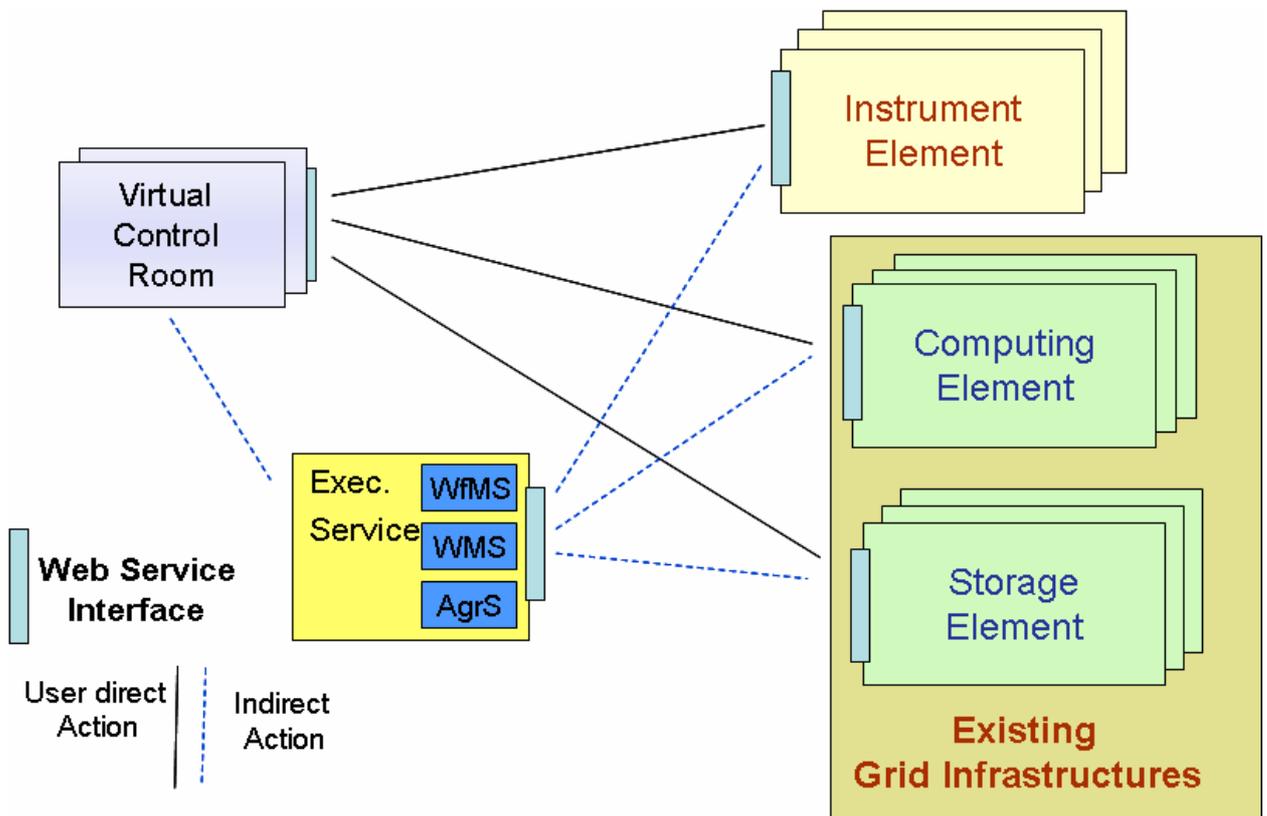

Figure 2. Interaction between the IE and other grid components.

Below the important IE characteristics are described [16].

The IE provides the following functional requirements:

- uniform model of an instrument;
- standard grid access to the instruments;
- capability to cooperate between different instruments belonging to different institutes and Virtual Organisations (VOs).

Remark: An IE user is not only a human being but also any application with certain rights specified in an electronic grid certificate.

An IE user can have one of the following roles shown in Figure 3 [16]:

1. Observer – can monitor the instrument;
2. Operator – can have an access to an instrument configuration, as well as the right to control and monitor the instrument;
3. Administrator – can create an instrument configuration which then becomes accessible for 1 and 2.

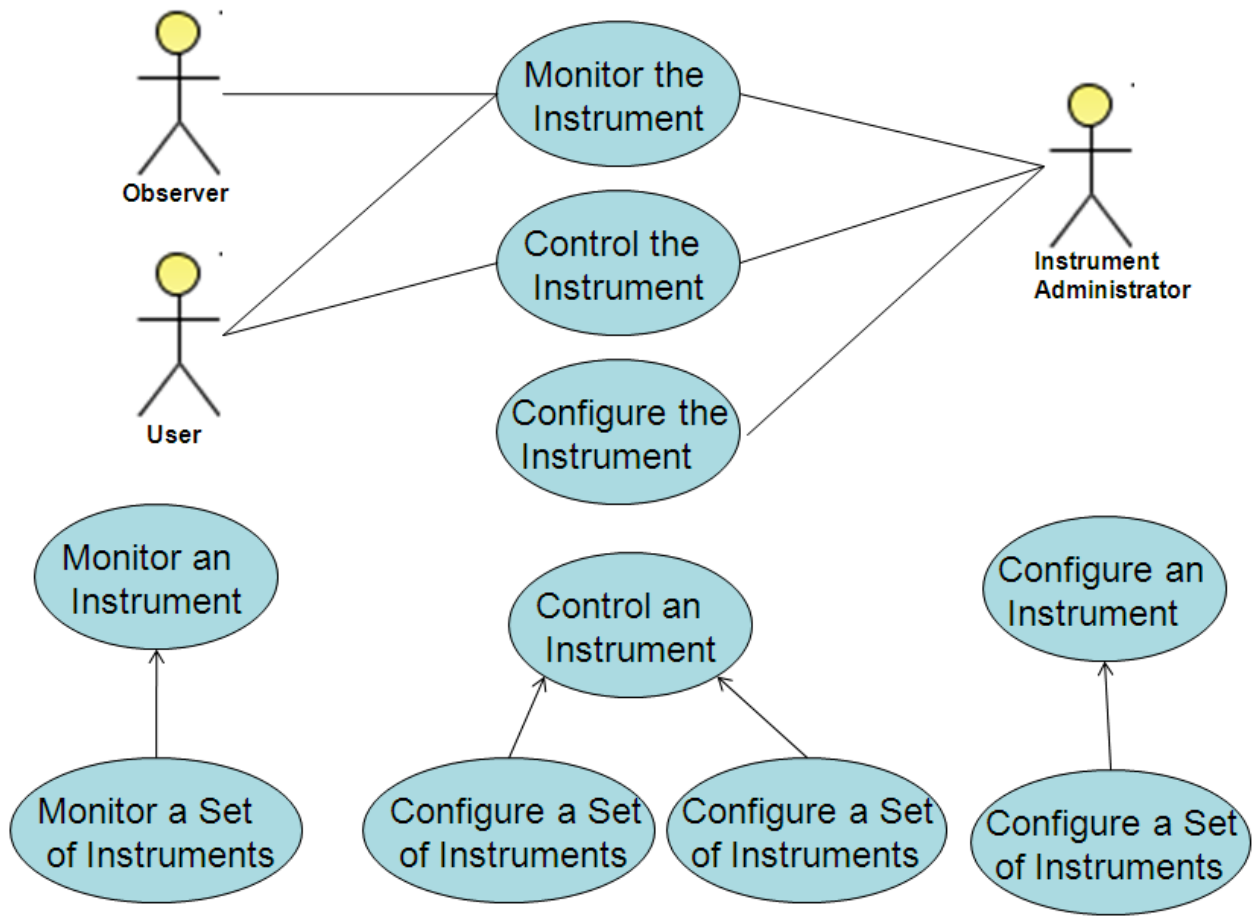

Figure 3. IE use cases.

At any moment there may be many observers and administrators, but only one operator which uses (i.e. controls) the instrument.

If the instrument is complex and has independent modules (as with the LHC CMS detector [11,12]) then several operators can use their separate module.

The IE was developed based on the functional (see above) and technical requirements (see Table 1) [16].

**Table 1** Nonfunctional requirements

| Nonfunctional requirements | Type |
|---|---|
| $O(10_4)$ nodes/instruments must be controlled and monitored | Scalability |
| The nodes/instrument should be accessed through the web. | Remote access |
| The nodes/instrument should be accessed in a homogeneous way. | Standardisation |
| Round-trip time to reach all the nodes must be in the order of human reaction time. | Quality of Service |
| Online diagnostics and possible error recovery | Autonomic |

An instrument or a set of instruments can be both very simple and very complex. Therefore, the following instrument categories were introduced when making an abstract model of the instrument:

- dummy instrument
- smart instrument
- smart instrument in an ad hoc network

Figure 4 gives a uniform model of instrument control [15].

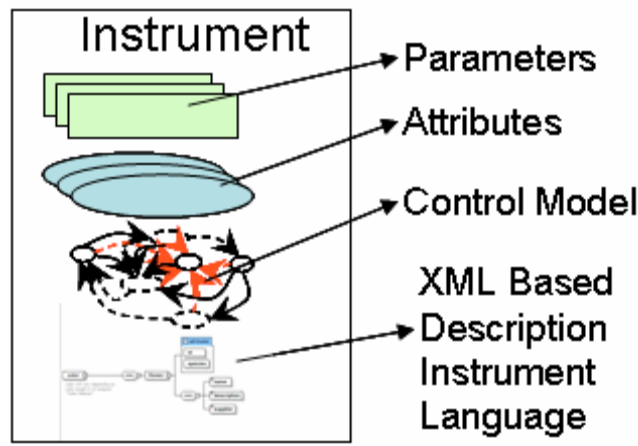

Figure 4. Uniform model of instrument control.

Figure 5 shows a flow-chart describing the instrument integration into the grid [16].

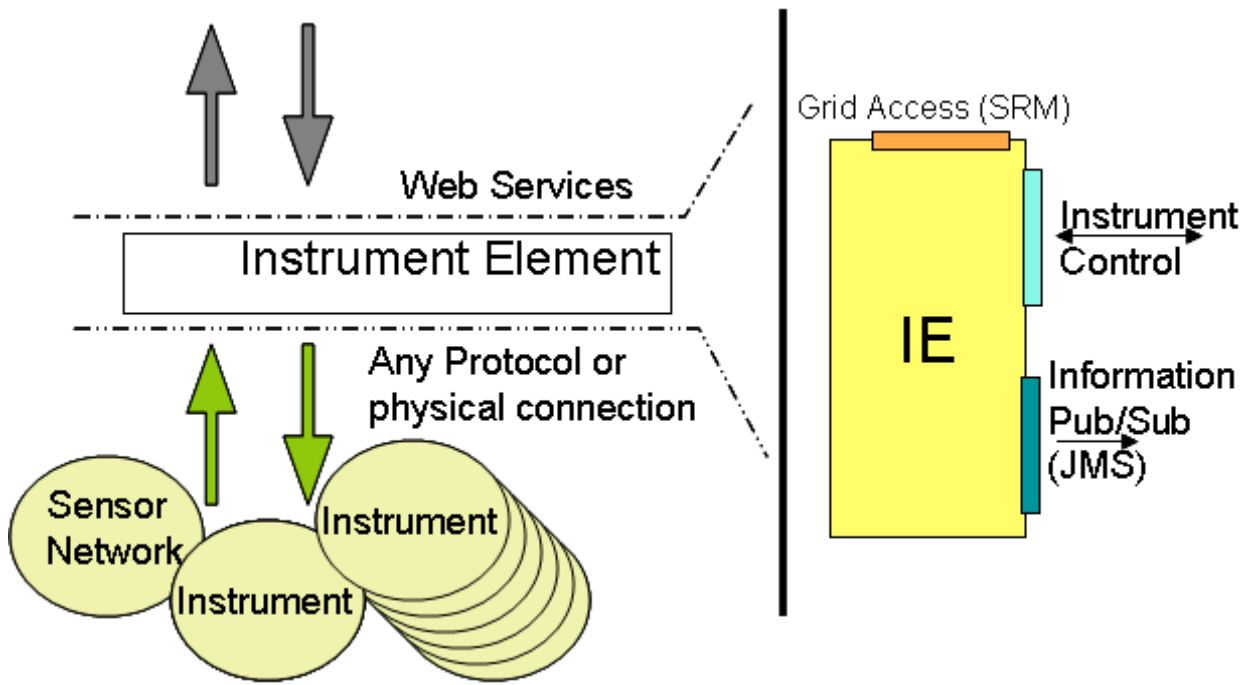

Figure 5. Instrument integration into the grid

**The Grid and Some Metrology Tasks**

Below metrology areas are described where the application of the grid could significantly improve the efficiency of the metrological activity.

1. *Commercial transactions with consumers*

Electricity, gas, heat, water meters, etc. are constituent elements of the system of commercial transactions with end consumers and are under *legal metrological control*.

There is little doubt that the grid as a global distributed computing system with security issues solved systematically and at the highest level of modern information technology will be introduced in the banking sector, and commercial transactions with end customers will be integrated into this system.

2. Legal metrology specifies the requirements for the *long-term storage and transfer of measurement data*, as well as for the uploading of new software versions for measuring instruments [7]. The document WELMEC 7.2 gives these requirements in the extensions L, T, D.

The Grid has specifically been designed to solve such tasks. It should be reminded that the data hypothetically demonstrating the existence of the Higgs boson in the LHC experiments are a negligible part in the whole array ~ 1PB per year of collision results [9,10]. Therefore a task was set to prevent any distortion or loss of experiment data.

Thus, the use of the grid and its Instrument Element component could solve the problems specified in WELMEC 7.2 extensions L, T, D.

3. *Remote comparisons of measurement standards, reference gauges, and devices* [4]
In many cases the modern computer technology allows remote comparisons of measuring instruments.

There are international programs in clock synchronisation, navigation and other areas within which remote comparisons are carried out.

The Grid can move these approaches to an absolutely new level both qualitatively and quantitatively providing better efficiency and data security.

4. *Key comparisons of measurement standards*
These comparisons are often time-consuming and expensive.

The Grid can provide effective tools to carry out remote comparisons of measurement standards, uniform calculations, and data transfer.

Most operations, sometimes even all of them, can be made remotely.

5. *Linux OS embedded instruments*
The software of measuring instruments becomes more and more complex. Linux OS embedded instruments are more widely used. As measuring instruments they are under legal metrological control. The issues of testing such systems and assuring the security of measurement data are crucial. The above mentioned L, T, D requirements should be met, which is a new and difficult task for these systems [5].

These tasks are though well-known to grid developers. In fact, LHC is the most complex measuring instrument in the history of our civilisation where similar and many other tasks are solved using the grid. The tools the grid provides as well as the newly developed IE give all opportunities to effectively and safely use measuring instruments with embedded Linux OS.

Another approach to solve measuring tasks with complex software is the *'thin client'* concept. In other words, instruments should have minimum software integrating them into the grid, and the complex processing of raw data should be performed in the grid environment.

The choice of an approach is up to measuring instrument developers.

6. *"Reference" software*
One of the main methods to study measuring instrument software is to compare it with reference software. However, the term 'reference software' is not sufficiently determined. Besides, the comparison can be made based on both the results of code execution and the literal code matching.

In the future, the Grid could assign a clear meaning to the reference software concept and make the comparison on a new, higher level. In our opinion, leading metrological organisations could set and complete the task of creating metrological reference software repository as one of grid resources similarly to the gLite middleware repositories [9].

In particular, this is the very software to be used in key comparisons of measurement standards.

A relatively small group of highly skilled metrologists, mathematicians, and programmers would be quite enough to create a reference program resource.

7. *Reference test data*
Besides the issue in item 6 there is also a problem to create a grid resource for reference test data [8] or, in a broader meaning, test tasks. Such tasks have been used for many years to test clusters within the computer experiment assurance system at the LHC based on the grid [11,12,13].

8. *Smart Electrical Grid*
Recently, much attention has been paid to the so-called 'smart meters' that allow energy to be saved. It is stated in [6] that one of the major goals of the European Metrology Research Programme (EMRP) is "to develop a metrological measurement infrastructure in Europe to support successful implementation of a Smart Electrical Grid".

Remark: Within the 'Smart Electrical Grid' concept the 'Grid' term means electric power mains, which differs from the meaning of Grid as an information computer network this paper is dealt with.

EMRP specifies the following research areas in Smart Electrical Grid [6]:
"a) Measurement framework for monitoring stability of smart grids via application of reliable and accurate Phasor Measurement units;
b) Traceable on-site energy measurement systems (smart meters) for ensuring fair energy trade;
c) Remote on-site measurement of power quality and efficiency; and
d) Modelling, simulation and network analysis of the system state of smart grids".

The available Grid concept realisation - the gLite middleware - together with the Instrument Element provides a complete set of tools to create applications for solving the above-mentioned problems in this area.

Taking into account the conceptuality and integrity of the computing resources management system and the information security system, as well as the genericity of the instrument model (see above) in the IE, the development of *grid-independent* applications duplicating the solutions for these problems seems inexpedient, both with regard to time, financial, and manpower costs and to the quality of solving the problems.

## Conclusions

A number of crucial metrological problems need to be adequately supported by the brand-new computer technologies. We are convinced that Grid is one of such technologies that meet the global challenges of today. The creation of the Instrument Element as a grid component makes this system to be an optimal environment for effective monitoring, management and servicing of measuring resources which has the highest level of measurement data transfer, storage and processing safety and reveals new opportunities to track measurement procedures and assure a high level of confidence to these measurements. In addition to the communities already using the Grid [17,18 ]:
- earthquake community,
- environment community,
- experimental science community,
the metrological community can obtain an extremely powerful tool to realise its goals.